\documentclass[12pt,aps,preprint]{revtex4}

\usepackage{amsmath}
\usepackage{mathrsfs}
\usepackage{bm}
\usepackage{graphicx}
\usepackage{subfigure}

\begin{document}

\begin{titlepage}	
\title{Backward waves in a grounded bilayer slab containing double-negative (DNG) and double-positive (DPS) metamaterials}

\author{E.~Cojocaru}

\affiliation{Department of Theoretical Physics \\ 
Horia Hulubei National Institute of Physics and Nuclear Engineering \\
P.O.B. MG-6, 077125 Magurele, Romania}

\email{ecojocaru@theory.nipne.ro} 

\begin{abstract}
Simple dispersion relations for the guided modes in a grounded DNG/DPS bilayer slab are given in terms of normalized parameters. Relations corresponding to the grounded single-layer DNG slab are refound as specific cases. Numerical examples are given showing dispersion curves of the lower order modes and the respective total normalized power carried on the propagation direction. Snapshots obtained by the finite-difference time-domain method are provided showing the electromagnetic field inside the grounded DNG/DPS bilayer slabs. Since an important characteristic of the guided modes in the slab containing a DNG layer is the existence of a turning point (TP) at which the power carried by each mode of order $m>0$ equals zero and changes the sign, we present implicit relations at the TP for the normalized parameters of the guided modes in the grounded DNG/DPS and DNG slabs. We show that a thin DPS layer coating on the grounded DNG slab produces a shift of the TP on the dispersion curve.
\end{abstract}
\maketitle
\end{titlepage}
\section{Introduction}
Recently, the materials having negative permittivity and negative permeability, which are labeled as double-negative (DNG) materials, have attracted much attention, especially because of their unusual physical properties, which are different from those of conventional double-positive (DPS) materials. Veselago \cite{1} was the first to study theoretically the DNG materials, but these materials became a topic of high interest after the first experimental demonstration of their negative refractive index \cite{2}. Much attention has been paid to their potential applications, a pioneering work being Pendry's suggestion of perfect lenses made of DNG materials \cite{3}. Characteristics of these materials can be found for example in \cite{4,5,6}. Electromagnetic wave propagation in these materials has been extensively investigated theoretically and experimentally \cite{7,8}. Theoretical studies have shown that such materials are capable of supporting backward waves \cite{9,10,11,12}. Waveguides containing DNG layers are described for example in \cite{12,13,14,15,16,17,18}. Guided waves in grounded single-layer slabs of DNG materials have been analyzed in \cite{12,19,20,21,22}.

In this paper we consider grounded slabs containing DNG/DPS bilayers. Characteristics of the single-layer DNG slabs are refound as specific cases from those for the bilayer slabs. Dispersion relations are given in terms of normalized parameters. Numerical examples show dispersion curves of the lower order modes and the total normalized power carried on the propagation direction by the respective modes. Examples of electromagnetic fields inside the bilayer slab are also given.  Snapshots obtained by the finite-difference time-domain (FDTD) method \cite{23} are provided. An important characteristic of the guided modes in the slab containing a DNG layer is the existence of a turning point (TP) at which the power carried by each mode of order $m>0$ equals zero and changes the sign. We present implicit relations for the normalized parameters of the guided modes in the grounded DNG/DPS and DNG slabs at the TP.  

\section{General relations}
\begin{figure}[ht]
\includegraphics[width=5cm]{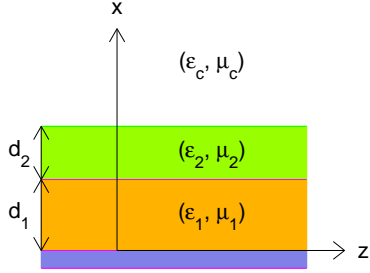}
\caption{\label{fig:fig_1}Geometry for the grounded bilayer slab of layer thicknesses $d_j$ and relative material constants $(\epsilon_j,\mu_j)$, where $j=1,2$, with a cover medium of relative material constants $(\epsilon_c,\mu_c)$.}
\end{figure}

The geometry of the grounded bilayer slab is shown in Fig.~\ref{fig:fig_1}. The slab of thickness $d_1+d_2$ is infinite along the $y$ and $z$ directions. It is grounded by a perfectly electric conducting substrate in the plane $x=0$. The relative material constants are $(\epsilon_j,\mu_j)$, with $j=1,2$, for the layers of the slab, $(\epsilon_c,\mu_c)$ for the cover medium, and $(\epsilon_s,\mu_s)$ for the substrate, so that $\epsilon_1\mu_1\geq\epsilon_2\mu_2>\epsilon_c\mu_c$. For simplicity, the analysis in this paper is restricted to the case of frequency-independent and lossless material constants in order to obtain relations between the geometrical and material parameters. Only in the FDTD numerical simulations we use the Drude-Lorentz model for the material constants of the DNG layer \cite{23}. We consider the modes with a time harmonic variation $\exp(i\omega t)$ propagating in the $z$ direction, their electromagnetic fields varying as $\exp{i(\omega t-\beta z)}$, where $\beta$ is the modal phase constant. For transversal magnetic (TM) modes, the magnetic field is along the $y$ direction, whereas it is the electric field along the $y$ direction in the case of transversal electric (TE) modes. 
The magnetic field $H_y$ for the TM modes takes the form
\begin{eqnarray}
\label{eq:1} 
0<x<d_1 \qquad &H_y=\cos(k_{t1}x), \nonumber \\
d_1<x<d_1+d_2 \qquad &H_y=A_1\cos[k_{t2}(x-d_1)]+B_1\sin[k_{t2}(x-d_1)],  \nonumber \\
x>d_1+d_2 \qquad &H_y=A_2\exp\{-\kappa_c[x-(d_1+d_2)]\},
\end{eqnarray}
where the factor $\exp(-i\beta z)$ is skipped, and
\begin{equation}
\label{eq:2}
A_1=\cos(k_{t1}d_1), \qquad B_1=-\frac{k_{t1}\epsilon_2}{k_{t2}\epsilon_1}\sin(k_{t1}d_1), \qquad
A_2=A_1\cos(k_{t2}d_2)+B_1\sin(k_{t2}d_2),
\end{equation}
\begin{eqnarray}
\label{eq:3}
k_{tj}=(k_0^2\epsilon_j\mu_j-\beta^2)^{1/2} \qquad \textrm{when}\quad\bar{\beta}^2<\epsilon_j\mu_j, \nonumber \\ 
k_{tj}=-i(\beta^2-k_0^2\epsilon_j\mu_j)^{1/2} \qquad \textrm{when}\quad\bar{\beta}^2>\epsilon_j\mu_j,    
\end{eqnarray}
with $j=1,2$, $\bar{\beta}=\beta/k_0$ is the relative phase constant, $k_0=\omega(\epsilon_0\mu_0)^{1/2}$ is the wavenumber in vacuum, and 
\begin{equation}
\label{eq:4}
\kappa_c=(\beta^2-k_0^2\epsilon_c\mu_c)^{1/2}.
\end{equation}
Similarly, one can express the electric field $E_y$ for the TE modes.
The total normalized power carried by each mode on the $z$ direction takes the form
\begin{equation}
\label{eq:5}
\bar{P}=(P_1+P_2+P_c)/(|P_1|+|P_2|+|P_c|),
\end{equation}
where $P_1$, $P_2$, and $P_c$ are given up to a common factor by 
\begin{equation}
\label{eq:6}
P_1\propto\frac{1}{\epsilon_1}\int_0^{d_1} |H_y|^2 \textrm{d}x, \qquad P_2\propto\frac{1}{\epsilon_2}\int_{d_1}^{d_1+d_2} |H_y|^2 \textrm{d}x, \qquad
P_c\propto\frac{1}{\epsilon_c}\int_{d_1+d_2}^{\infty} |H_y|^2 \textrm{d}x,
\end{equation}
for the TM modes. Similarly, one can express $P_1$, $P_2$, and $P_c$ for the TE modes.

\section{Dispersion relations} 
Dispersion relations are determined either from the poles of the complex-amplitude reflection coefficients expressed in terms of the transfer matrix elements \cite{18,24,25} taking into account that $\epsilon_s=-\infty$ and $\mu_s=0$ \cite{26}, or by applying the boundary conditions of the tangential electromagnetic fields at interfaces. One obtains the following dispersion relation
\begin{equation}
\label{eq:7}
\frac{\epsilon_ck_{t1}}{\epsilon_1\kappa_c}\tan(k_{t1}d_1)=\left[1-\frac{\epsilon_ck_{t2}}{\epsilon_2\kappa_c}\tan(k_{t2}d_2)\right]/\left[1+\frac{\epsilon_2\kappa_c}{\epsilon_ck_{t2}}\tan(k_{t2}d_2)\right],
\end{equation}
for the TM modes, and
\begin{equation}
\label{eq:8}
\frac{\mu_1\kappa_c}{\mu_ck_{t1}}\tan(k_{t1}d_1)=\left[1+\frac{\mu_2\kappa_c}{\mu_ck_{t2}}\tan(k_{t2}d_2)\right]/\left[\frac{\mu_ck_{t2}}{\mu_2\kappa_c}\tan(k_{t2}d_2)-1\right],
\end{equation}
for the TE modes, where $k_{tj}$, with $j=1,2$, and $\kappa_c$ are given by relations (\ref{eq:3}) and (\ref{eq:4}), respectively. In the case $\bar{\beta}^2<\epsilon_2\mu_2<\epsilon_1\mu_1$ we introduce the normalized parameters 
\begin{eqnarray}
\label{eq:9}
\widetilde{u}_c=(\bar{\beta}^2-\epsilon_c\mu_c)^{1/2}, \qquad  u_j=(\epsilon_j\mu_j-\bar{\beta}^2)^{1/2},\qquad 
v_j=k_0d_j, \qquad j=1,2, \nonumber \\
\gamma_j=\epsilon_j\widetilde{u}_c/(\epsilon_c u_j) \quad \textrm{for~~TM~~modes}, \qquad
\gamma_j=-\mu_c u_j/(\mu_j\widetilde{u}_c) \quad \textrm{for~~TE~~modes}.  
\end{eqnarray}
Using notations (\ref{eq:9}) in (\ref{eq:7}) and (\ref{eq:8}) gives
\begin{equation}
\label{eq:10}
\frac{\tan(u_1v_1)}{\gamma_1}=\frac{1-\tan(u_2v_2)/\gamma_2}{1+\gamma_2\tan(u_2v_2)}.
\end{equation}
Thus, the TM and TE modes satisfy the normalized dispersion relation  
\begin{equation}
\label{eq:11}
v_1=\frac{1}{u_1}\left[\tan^{-1}\frac{\gamma_1[1-\tan(u_2v_2)/\gamma_2]}{1+\gamma_2\tan(u_2v_2)}+m\pi\right], \quad m=0,1,2,\dots
\end{equation}
When $d_2=0$, then $v_2=0$, and (\ref{eq:11}) reduces to the simple form
\begin{equation}
\label{eq:12}
v_1=(1/u_1)(\tan^{-1}\gamma_1+l\pi), \qquad l=0,1,2,\dots
\end{equation}
for the TM and TE modes, the cutoffs $\bar{\beta}=(\epsilon_c\mu_c)^{1/2}$ being at $v_1=l\pi$, with $l=0,1,2,\dots$, for the TM modes, and $v_1=(2l-1)\pi/2$, with $l=1,2,\dots$, for the TE modes \cite{19}. Note that, for simplicity, we label the modes by $l$ or $m$ without any reference to the standard mode labeling of DPS slabs. In the case $\epsilon_2\mu_2<\bar{\beta}^2<\epsilon_1\mu_1$, tacking into account relation (\ref{eq:3}), we denote
\begin{eqnarray}
\label{eq:13}
u_2=-i\widetilde{u}_2, \qquad \widetilde{u}_2=(\bar{\beta}^2-\epsilon_2\mu_2)^{1/2}, \qquad
\gamma_2=i\widetilde{\gamma}_2, \nonumber \\
\widetilde{\gamma}_2=\epsilon_2\widetilde{u}_c/(\epsilon_c \widetilde{u}_2) \quad \textrm{for~~TM~~modes}, \qquad
\widetilde{\gamma}_2=\mu_c\widetilde{u}_2/(\mu_2\widetilde{u}_c) \quad \textrm{for~~TE~~modes}.  
\end{eqnarray}
Then, the TM and TE modes satisfy the normalized dispersion relation
\begin{equation}
\label{eq:14}
v_1=\frac{1}{u_1}\left[\tan^{-1}\frac{\gamma_1[1+\tanh(\widetilde{u}_2v_2)/\widetilde{\gamma}_2]}{1+\widetilde{\gamma}_2\tanh(\widetilde{u}_2v_2)}+m\pi\right].
\end{equation}
In the case $\bar{\beta}^2>\epsilon_1\mu_1>\epsilon_2\mu_2$ both $\gamma_1$ and $\gamma_2$ have imaginary values, but there are real solutions to $\bar{\beta}$ of the $\textrm{TE}_0$ mode \cite{18}.  We denote $\widetilde{u}_1$ and $\widetilde{\gamma}_1$ like in (\ref{eq:13}). Then, the evanescent mode $\textrm{TE}_0$ of the grounded DNG/DPS bilayer slab satisfies the normalized dispersion relation
\begin{equation}
\label{eq:15}
v_1=\frac{1}{\widetilde{u}_1}\tanh^{-1}\left[-\frac{\widetilde{\gamma}_1[1+\tanh(\widetilde{u}_2v_2)/\widetilde{\gamma}_2]}{1+\widetilde{\gamma}_2\tanh(\widetilde{u}_2v_2)}\right].
\end{equation}
Since $\widetilde{\gamma}_1<0$, $v_1$ in (\ref{eq:15}) has real values, with $v_1\rightarrow 0$ when $\bar{\beta}\rightarrow\infty$. When $v_2=0$, relation (\ref{eq:15}) becomes
\begin{equation}
\label{eq:16}
v_1=(1/\widetilde{u}_1)\tanh^{-1}(-\widetilde{\gamma}_1).
\end{equation}

\section{Numerical examples}
\begin{figure}[ht]
\subfigure[]{\includegraphics[]{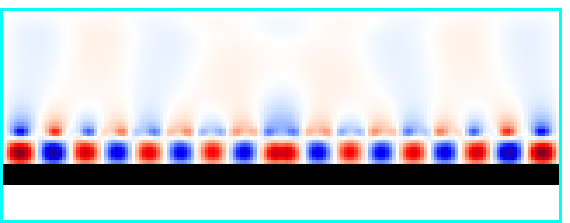}}\qquad
\subfigure[]{\includegraphics[]{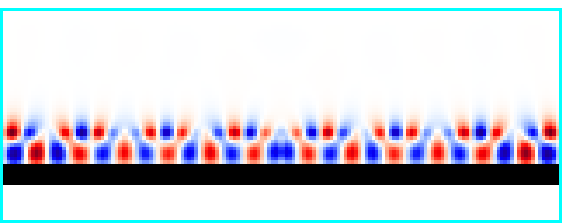}}
\caption{\label{fig:fig_4} Snapshots of TE backward propagating waves (a) in the grounded lossless DNG slab of frequency-dependent material constants $(\epsilon_1,\mu_1)=(-4,-2)$ with $v_1=1.8$ and $\omega/(2\pi c)=0.28$, and (b) in the grounded lossless DNG/DPS slab of frequency-dependent material constants $(\epsilon_1,\mu_1)=(-4,-2)$ and frequency-independent material constants $(\epsilon_2,\mu_2)=(2,1.5)$, with $v_1=2.25$, $d_2=d_1/2$, and $\omega/(2\pi c)=0.36$. The sinusoidal continuous source is placed at the center of the DNG layer.}
\end{figure}

For simplicity we consider the cover medium is air $(\epsilon_c=\mu_c=1)$. 
\begin{figure}[ht]
\includegraphics[width=9cm]{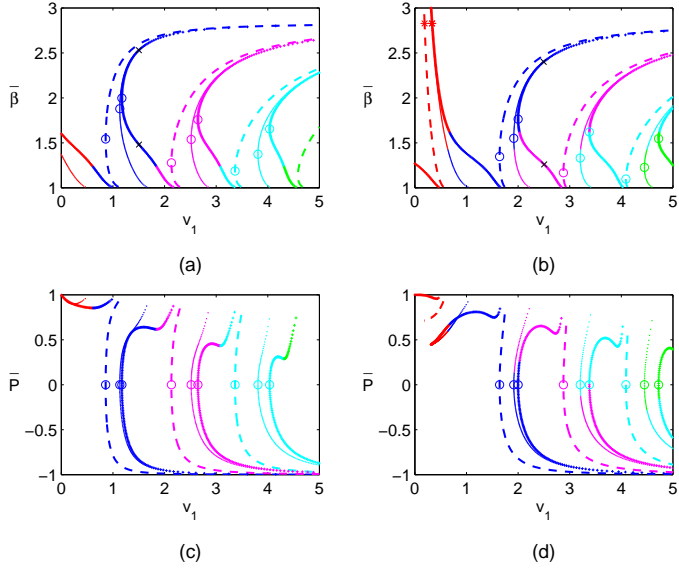}
\caption{\label{fig:fig_2} Dispersion curves for (a) TM and (b) TE modes in a grounded DNG/DPS bilayer slab of material constants $(\epsilon_1,\mu_1)=(-4,-2)$ and $(\epsilon_2,\mu_2)=(2,1.5)$, when $v_2$ is constant: $v_2=0$ (dashed line), $v_2=1$ (thin line, marker .), and $v_2=2$ (thick line, marker +).  Only the lower order modes are considered: $m=0$ (red), $m=1$ (blue), $m=2$ (magenta), $m=3$ (cyan), and $m=4$ (green). The TP of each mode is marked by a small circle.  The total normalized power $\bar{P}$ carried by each mode on the propagation direction is represented in (c) and (d) for the TM and TE modes, respectively.}
\end{figure}
\begin{figure}[ht]
\includegraphics[width=9cm]{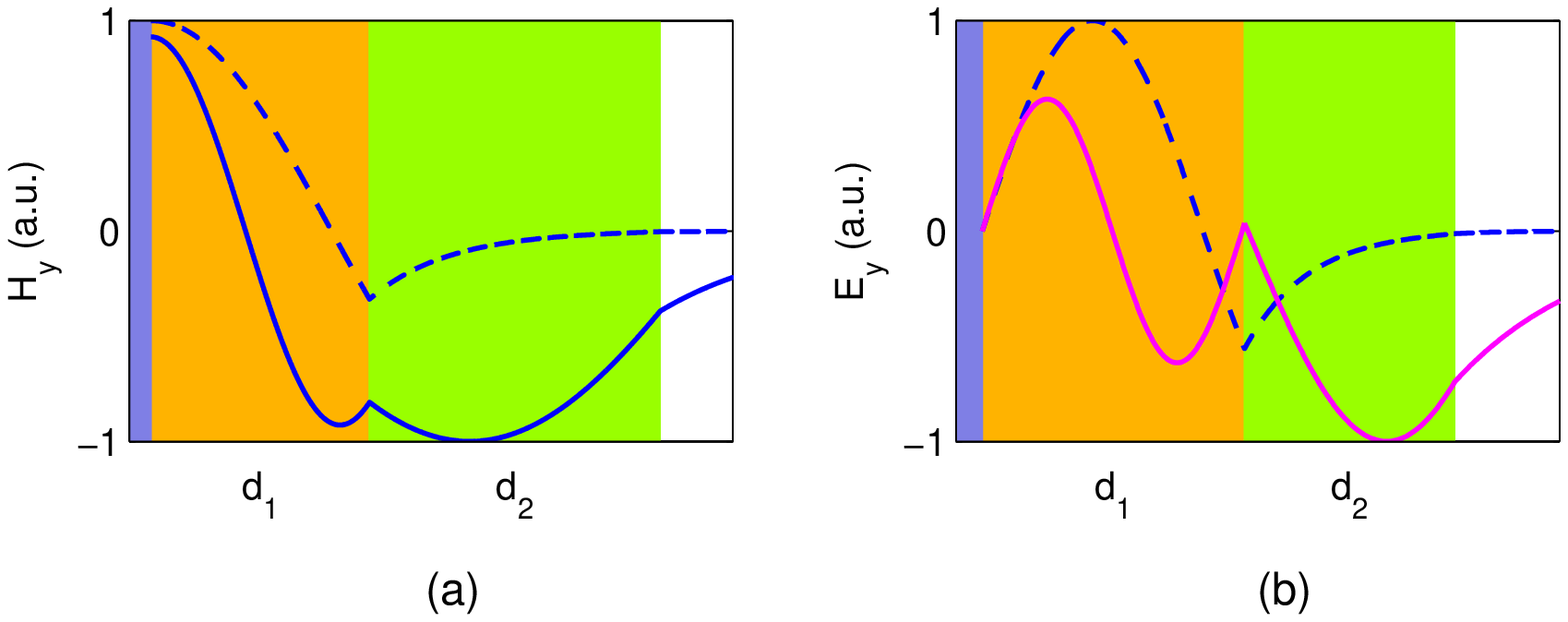}
\caption{\label{fig:fig_3} The fields $H_y$ in (a) and $E_y$ in (b) for the TM and TE modes of a grounded DNG/DPS bilayer slab at the points shown in Figs.~\ref{fig:fig_2}(a) and (b) by black markers, when $v_2=2$. For the TM mode in (a), $v_1=1.5$, with $\bar{\beta}=1.48$ (full line), and $\bar{\beta}=2.53$ (dashed line). For the TE mode in (b), $v_1=2.5$, with $\bar{\beta}=1.26$ (full line), and $\bar{\beta}=2.4$ (dashed line).}
\end{figure} 
Dispersion curves are shown in Figs.~\ref{fig:fig_2}(a) and (b) for the TM and TE modes, respectively, when $(\epsilon_1,\mu_1)=(-4,-2)$ and $(\epsilon_2,\mu_2)=(2,1.5)$, at $v_2=0,1,~\textrm{and}~2$. The dispersion curves overlap in the upper region, at $\bar{\beta}^2>\epsilon_2\mu_2$, but in the lower region, at $\bar{\beta}^2<\epsilon_2\mu_2$, there are composed modes \cite{16}. Thus, it is more easy to look after the modes starting from the upper region of the dispersion curves, at $\bar{\beta}^2>\epsilon_2\mu_2$. In Fig.~\ref{fig:fig_2}(b), the $\textrm{TE}_0$ mode allows real solutions at $\bar{\beta}^2\geq\epsilon_1\mu_1$, the starting point being marked by an asterisk. For each mode of $m\neq0$, the total power carried on the $z$ direction equals zero at the TP, as shown in Figs.~\ref{fig:fig_2}(c) and (d). Note that the TPs are more distinctly seen on the $\bar{P}$ against $v_1$ curves than on the dispersion curves. We denote the values of $\bar{\beta}$ and $v_1$ at the TP by $\bar{\beta}_{\textrm{TP}}$ and $v_{1_{\textrm{TP}}}$, respectively. At $\bar{\beta}<\bar{\beta}_{\textrm{TP}}$, the total power $\bar{P}$ is positive (forward wave), whereas at $\bar{\beta}>\bar{\beta}_{\textrm{TP}}$, the total power carried by each mode of $m\neq0$ is negative, (backward wave) \cite{11,12,17,18}. In Fig.~\ref{fig:fig_2}, one can see that, the greater is $v_2$, the greater are $\bar{\beta}_{\textrm{TP}}$ and $v_{1_{\textrm{TP}}}$. At given $v_2$, the greater the order of the mode, the smaller $\bar{\beta}_{\textrm{TP}}$ and the greater $v_{1_{\textrm{TP}}}$. The fields $H_y$ and $E_y$ in the slab are shown in Figs.~\ref{fig:fig_3}(a) and (b) for the TM and TE modes, respectively. At $\bar{\beta}^2<\epsilon_2\mu_2$ the field is maximum in the DPS layer, while at $\bar{\beta}^2>\epsilon_2\mu_2$ the field is maximum in the DNG layer. Snapshots of TE backward propagating waves obtained by the FDTD method \cite{23} are shown in Figs.~\ref{fig:fig_4}(a) and (b) for the grounded DNG and DNG/DPS slabs, respectively. The Drude-Lorentz model was used for the dispersive material constants of the lossless DNG layer \cite{23}, whereas frequency-independent and lossless material constants were considered for the DPS layer. The computation space is $160\times60$ cells and 200 time units is approximately the time to run a simulation. A perfectly matched layer was placed at the upper side of the simulation domain. The slowly rumped continuous sinusoidal source is placed at the center of the DNG layer. We considered $v_1=1.8$ with $\omega/(2\pi c)=0.28$ in Fig.~\ref{fig:fig_4}(a), and $v_1=2.25$, $d_2=d_1/2$, and $\omega/(2\pi c)=0.36$ in Fig.~\ref{fig:fig_4}(b). Contrary to the forward propagating waves in a classical slab, the backward waves are propagating from the lateral sides toward the centered source \cite{12}. 
\begin{figure}[ht]
\includegraphics[width=9cm]{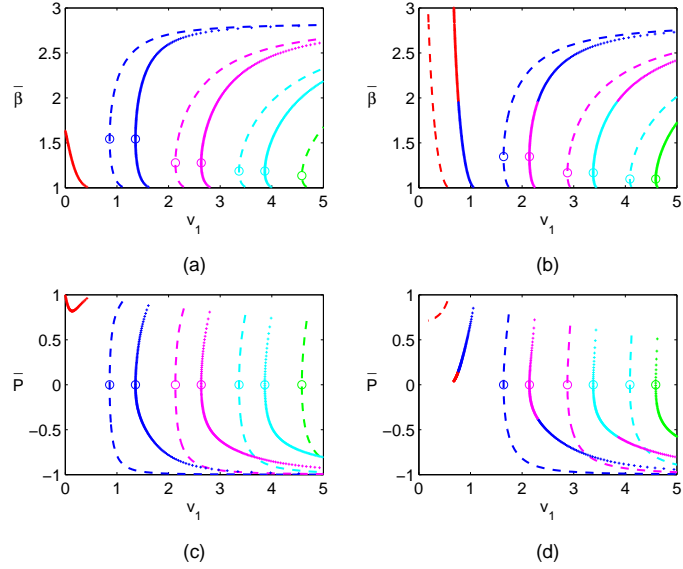}
\caption{\label{fig:fig_5}Dispersion curves for (a) TM and (b) TE modes in a grounded DNG/DPS bilayer slab of material constants $(\epsilon_1,\mu_1)=(-4,-2)$ and $(\epsilon_2,\mu_2)=(4,2)$, when $v_2=0$ (dashed line)and $v_2=0.5$ (thick line, marker +). The colors are kept the same like in Fig.~\ref{fig:fig_2}. The TP of each mode is marked by a small circle.  The total normalized power $\bar{P}$ carried by each mode on the propagation direction is represented in (c) and (d) for the TM and TE modes, respectively.}
\end{figure} 

Now, consider a grounded DNG/DPS bilayer slab of relative material constants $(\epsilon_1,\mu_1)=(-4,-2)$ and $(\epsilon_2,\mu_2)=(4,2)$. Note that $d_1\neq d_2$. Dispersion curves are shown in Figs.~\ref{fig:fig_5}(a) and (b) for the TM and TE modes, respectively, at $v_2=0$ and $0.5$. The total normalized power $\bar{P}$ carried by each mode on the propagation direction is represented in (c) and (d) for the TM and TE modes, respectively. Dispersion curves at $v_2\neq0$ are translated as against those at $v_2=0$, so that, for a given mode of $m\neq0$, 
\begin{equation}
\label{eq:17}
\bar{\beta}_{_{\textrm{TP}}}\Big|_{v_2\neq0}=\bar{\beta}_{_{\textrm{TP}}}\Big|_{v_2=0}, \qquad
v_{1_{_{\textrm{TP}}}}\Big|_{v_2\neq0}=v_{1_{_{\textrm{TP}}}}\Big|_{v_2=0}+v_2.
\end{equation}
\begin{figure}[ht]
\includegraphics[width=9cm]{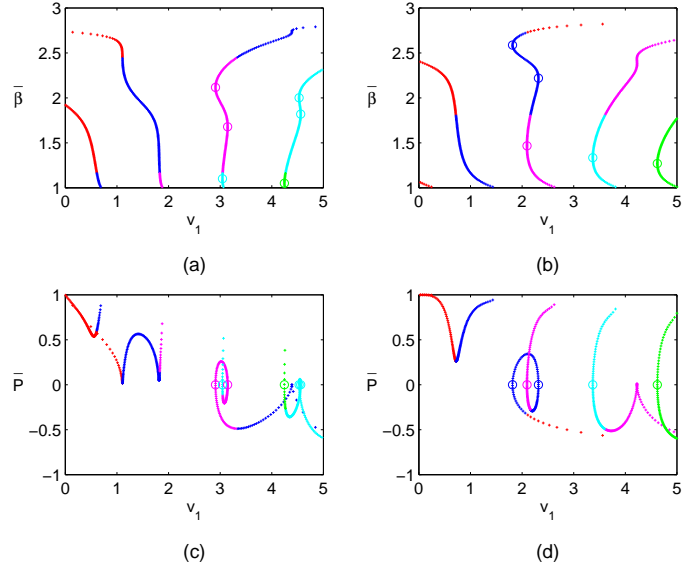}
\caption{\label{fig:fig_6}Dispersion curves for (a) TM and (b) TE modes in a grounded DNG/DPS bilayer slab of material constants $(\epsilon_1,\mu_1)=(-4,-2)$ and $(\epsilon_2,\mu_2)=(2,4)$, when $v_2=2$. The colors are kept the same like in Fig.~\ref{fig:fig_2}. The TPs are marked by small circles. The total normalized power $\bar{P}$ carried by each mode on the propagation direction is represented in (c) and (d) for the TM and TE modes, respectively.}
\end{figure}
\begin{figure}[ht]
\includegraphics[width=5cm]{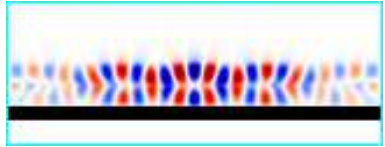}
\caption{\label{fig:fig_7} Snapshot of the TE backward propagating wave in the grounded lossless DNG/DPS slab of frequency-dependent material constants $(\epsilon_1,\mu_1)=(-4,-2)$ and frequency-independent material constants $(\epsilon_2,\mu_2)=(2,4)$, with $v_1=2$, $d_2=d_1$, and $\omega/(2\pi c)=0.32$. The sinusoidal continuous source is placed at the center of the DNG layer.}
\end{figure}
 
Now, consider a grounded DNG/DPS bilayer slab of relative material constants $(\epsilon_1,\mu_1)=(-4,-2)$ and $(\epsilon_2,\mu_2)=(2,4)$. Dispersion curves are shown in Figs.~\ref{fig:fig_6}(a) and (b) for the TM and TE modes, respectively, at $v_2=2$. The total normalized power $\bar{P}$ carried by each mode on the propagation direction is represented in (c) and (d) for the TM and TE modes, respectively. Intricate modes exist at $v_1>v_2$, as for example, the first mode at $v_1>v_2$ has three TPs. The total normalized power $\bar{P}$ has positive values between the two TPs of the same order $m$, as shown in Figs.~\ref{fig:fig_6}(c) and (d). The modes at $v_1>v_2$ have $\bar{P}<0$, whereas the modes at $v_1<v_2$ have $\bar{P}>0$. In Fig.~\ref{fig:fig_6} (d), the mode of $m=2$ has $\bar{P}\approx 0$ at $v_1=4.2282$, but $\bar{P}$ does not change the sign, and so, there is not a TP at that value of $v_1$. Snapshot of the TE backward propagating wave obtained by the FDTD method \cite{23} is shown in Fig.~\ref{fig:fig_7}, when $v_1=v_2=2$, with $\omega/(2\pi c)=0.32$. After a long time, more than 300 time units, the field stays concentrated near the continuous centered source.  

\section{Implicit relations at the TP}
Since the TP is an important characteristic of the modes in the grounded DNG/DPS and DNG slabs, here we present implicit relations for determining $\bar{\beta}_{\textrm{TP}}$ when $v_2$ is constant.  
\subsection{The case $\bar{\beta}^2<\epsilon_2\mu_2<\epsilon_1\mu_1$}
At the TP, $v_1$ is minimum and $\textrm{d}v_1/\textrm{d}\bar{\beta}=0$. Thus, using (\ref{eq:11}) gives the following implicit relation
\begin{equation}
\label{eq:18}
\frac{\gamma_1(1-t_2/\gamma_2)}{1+\gamma_2t_2}=\tan\left[\frac{\sigma\gamma_1F}{(1+\gamma^2t_2)^2+
\gamma_1^2(1-t_2/\gamma_2)^2}\right],
\end{equation}
where $\gamma_j$, with $j=1,2$, is defined by relation (\ref{eq:9}), $t_2=\tan(u_2v_2)$, 
\begin{equation}
\label{eq:19}
\sigma=-1 \quad \textrm{for~~TM~~modes}, \qquad \sigma=1 \quad \textrm{for~~TE~~modes},
\end{equation}
\begin{equation}
\label{eq:20}
F=\frac{\eta_{1c}}{\widetilde{u}_c^2}(1-t_2^2)-\frac{\eta_{12}t_2}{u_2^2}(\gamma_2-\frac{1}{\gamma_2})-
\sigma\frac{u_1^2v_2}{u_2}(1+t_2^2)(\gamma_2+\frac{1}{\gamma_2})+\frac{2\eta_{2c}u_1^2t_2^2}{u_2^2
\widetilde{u}_c^2},
\end{equation}
with $\widetilde{u}_c$ defined by (\ref{eq:9}) and
\begin{equation}
\label{eq:21}
\eta_{12}=\epsilon_1\mu_1-\epsilon_2\mu_2, \qquad \eta_{jc}=\epsilon_j\mu_j-\epsilon_c\mu_c,\qquad j=1,2.
\end{equation}
Once $\bar{\beta}_{\textrm{TP}}$ is determined with (\ref{eq:18}), the respective value $v_{1_{\textrm{TP}}}$ is obtained with (\ref{eq:11}). When $v_2=0$, the implicit relation (\ref{eq:18}) becomes
\begin{equation}
\label{eq:22}
\gamma_1=\tan\left\{\sigma\gamma_1\eta_{1c}/[\widetilde{u}_c^2(1+\gamma_1^2)]\right\}.
\end{equation}
This is the implicit relations for determining $\bar{\beta}_{\textrm{TP}}$ in the case of a grounded single-layer DNG slab.

\subsection{The case $\epsilon_2\mu_2<\bar{\beta}^2<\epsilon_1\mu_1$}
Using $\textrm{d}v_1/\textrm{d}\bar{\beta}=0$ in (\ref{eq:14}) gives the following implicit relation for determining $\bar{\beta}_{\textrm{TP}}$ when $v_2$ is constant,
\begin{equation}
\label{eq:23}
\frac{\gamma_1(1+\widetilde{t}_2/\widetilde{\gamma}_2)}{1+\widetilde{\gamma}_2\widetilde{t}_2}=\tan\left[\frac{\sigma\gamma_1
\widetilde{F}}{(1+\widetilde{\gamma}^2\widetilde{t}_2)^2+
\gamma_1^2(1+\widetilde{t}_2/\widetilde{\gamma}_2)^2}\right],
\end{equation}
where $\widetilde{\gamma}_2$ is defined by relation (\ref{eq:13}), $\widetilde{t}_2=\tanh(\widetilde{u}_2v_2)$, and
\begin{equation}
\label{eq:24}
\widetilde{F}=\frac{\eta_{1c}}{\widetilde{u}_c^2}(1+\widetilde{t}_2^2)+\frac{\eta_{12}\widetilde{t}_2}{\widetilde{u}_2^2}(\widetilde{\gamma}_2+\frac{1}{\widetilde{\gamma}_2})+
\sigma\frac{u_1^2v_2}{\widetilde{u}_2}(1-\widetilde{t}_2^2)(\widetilde{\gamma}_2-\frac{1}{\widetilde{\gamma}_2})+\frac{2\eta_{2c}u_1^2\widetilde{t}_2^2}{\widetilde{u}_2^2\widetilde{u}_c^2}.
\end{equation}
Once $\bar{\beta}_{\textrm{TP}}$ is determined with (\ref{eq:23}), the respective value $v_{1_{\textrm{TP}}}$ is obtained with (\ref{eq:14}).

\subsection{Numerical examples}
Using implicit relations (\ref{eq:18}) and (\ref{eq:23}) gives at $v_2=1$ in Fig.~\ref{fig:fig_2}(a) the following values of $[\bar{\beta}_{\textrm{TP}},v_{1_{\textrm{TP}}}]$: $[1.8782,1.1329]$ at $m=1$, $[1.5375,2.5163]$ at $m=2$, $[1.3741,3.8101]$ at $m=3$, and in Fig.~\ref{fig:fig_2}(b), the following values:  $[1.5496,1.9174]$ at $m=1$, $[1.3311,3.2039]$ at $m=3$, and $[1.2266,4.4481]$ at $m=4$. In Fig.~\ref{fig:fig_6}(a), we obtain for the first intricate mode at $v_1>v_2$ the following values of $[\bar{\beta}_{\textrm{TP}},v_{1_{\textrm{TP}}}]$: $[2.1159,2.9105]$ and $[1.6777,3.1455]$ at $m=2$, and $[1.1011,3.0452]$ at $m=3$, whereas in Fig.~\ref{fig:fig_6}(b), we obtain for the first intricate mode at $v_1>v_2$: $[2.5865,1.8154]$ and $[2.2191,2.3176]$ at $m=1$, and $[1.4665,2.0952]$ at $m=2$.
\begin{figure}[h]
\includegraphics[width=9cm]{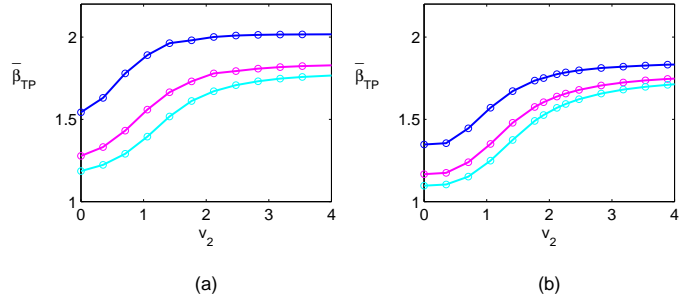}
\caption{\label{fig:fig_8}$\bar{\beta}_{\textrm{TP}}$ versus $v_2$ for (a) TM and (b) TE modes of a grounded DNG/DPS bilayer slab of relative material constants like in Fig.~\ref{fig:fig_2}.}
\end{figure} 
The values of $\bar{\beta}_{\textrm{TP}}$ are plotted versus $v_2$ in Fig.~\ref{fig:fig_8} for the TM and TE modes in a grounded DNG/DPS bilayer slab of relative material constants like in Fig.~\ref{fig:fig_2}. One can see that $\bar{\beta}_{\textrm{TP}}$ has a stronger increase at smaller values of $v_2$, whereas at greater values of $v_2$, it is practically independent of $v_2$. Thus, by coating a thin DPS layer on a grounded single-layer DNG slab, the TP is shifted towards greater values of $\bar{\beta}$ and $v_1$ for each TM or TE mode of order $m\neq0$, the shift being stronger for the TM modes in our numerical example.

\section{Conclusion}
In this paper we analyzed the TM and TE modes in a grounded slab containing DNG/DPS bilayers. Simple normalized dispersion relations were given for the guided and evanescent modes. Relations corresponding to the grounded single-layer DNG slab were refound as specific cases. Numerical examples were given showing dispersion curves of the lower order modes and the total normalized power carried on the propagation direction by the respective modes. Examples of electromagnetic fields inside the grounded DNG/DPS bilayer slabs were also given. Snapshots obtained by the FDTD method \cite{23}. Implicit relations for the normalized parameters at the TP were given for the grounded DNG/DPS and DNG slabs. We showed that a given mode could have several TPs, as for example, in Fig.~\ref{fig:fig_6} there are modes with three TPs, the plots of the total normalized power carried by each mode on the propagation direction being very useful in the designation of the TPs. Interesting results were obtained for the grounded DNG/DPS bilayer slabs of material constants $(\epsilon_1,\mu_1)=(-\epsilon_2,-\mu_2)$ [see Fig.~\ref{fig:fig_5} and relation (\ref{eq:17})] and of material constants $(\epsilon_1,\mu_1)=(-\mu_2,-\epsilon_2)$ (see Figs.~\ref{fig:fig_6} and \ref{fig:fig_7}). For simplicity, we considered $\epsilon_1\mu_1\geq\epsilon_2\mu_2$ and several examples were given, but relations presented in terms of normalized parameters in this paper could be applied also to other combinations for the relative material constants of the DPS and DNG materials. The analysis in this paper was restricted to the simplest case of lossless material constants. When absorption is necessarily taken into account, then the predicted behavior can qualitatively change.

\end{document}